\documentclass{article}
\usepackage[T1]{fontenc}
\usepackage[utf8]{inputenc}
\usepackage{ismir} 
\renewcommand{\permission}{}
\usepackage{amsmath,cite,url}
\usepackage{graphicx}
\usepackage{amssymb}
\usepackage{color}
\usepackage{multirow}

\usepackage[utf8]{inputenc}
\usepackage{CJKutf8}

\title{MeloBottleneck: Self-Supervised Melody Skeleton Extraction with a Latent Subsequence Bottleneck}

\multauthor
  {Fan Bu$^1$
   \hspace{0.8cm}
   Rongfeng Li$^{1,\ast}$
   \hspace{0.8cm}
   Linfeng Fan$^2$}
  {
   $^1$ Beijing University of Posts and Telecommunications\\
   $^2$ Central Conservatory of Music\\
   {\tt\small
   m.july@qq.com,
   lirongfeng@bupt.edu.cn,
   linfeng@ccom.edu.cn}\\
   {\small
   $^\ast$ Corresponding author:
   lirongfeng@bupt.edu.cn}
  }

\def\authorname{F. Bu, R. Li, and L. Fan}

\begin{document}

\maketitle
\thispagestyle{empty}

\begin{abstract}
Melody skeleton extraction aims to derive a shorter melody that preserves structural notes while removing ornaments. Prior methods rely on hand-crafted reduction rules or note-wise salience classifiers trained with heuristically or procedurally generated pseudo-labels. Such supervision can inherit generator bias and does not explicitly optimize a coherent reduced melody. We introduce MeloBottleneck, a self-supervised framework that represents a skeleton as a length-controlled, order-preserving latent subsequence. A hard-bottleneck extractor selects note events, a rhythmic-closure operator produces a self-consistent skeleton, and a re-ornamentation decoder reconstructs the input melody. Training combines reconstruction, a frozen autoregressive melody prior, ornament-invariant consistency across procedurally ornamented views, and ornament exclusion. We evaluate three regimes: synthetic out-of-distribution ornament-to-skeleton, TAVERN variation-to-theme, and Jiugong ornamented-to-gongche. A matched pseudo-label classifier excels on the synthetic benchmark, while MeloBottleneck transfers better, achieving competitive selection quality on TAVERN and Jiugong. Skeletonized melodies also improve BM25-based fragment retrieval, boosting Recall@K and MRR while reducing query time. Overall, the results suggest that learning skeletons as latent subsequences yields more robust transfer than pseudo-label imitation.
\end{abstract}

\section{Introduction}\label{sec:intro}

In monophonic symbolic music, melody skeleton extraction aims to derive a shorter melody that preserves structural notes while removing ornaments. Such skeletons can support symbolic comparison, retrieval, and theme tracing across variants, all tasks where surface elaboration can obscure shared melodic material. A practical extractor should therefore do more than identify locally important notes: it should retain information, produce a standalone reduced melody, remain stable under ornamentation, and allow explicit control over reduction length.

Existing approaches only partially meet these requirements. Music-theoretic and heuristic reducers are often interpretable, but they encode strong analytical priors that may be style-dependent \cite{Book83Lerdahl:01,JNMR06Hamanaka:01,ISMIR25Wang:01}. More recent learning-based work moves closer to note-wise structural prediction \cite{FDMC21Hu:01}. More generally, when skeleton extraction is trained from heuristic or procedural note-wise keep/delete pseudo-labels, the model can inherit generator bias. Moreover, note-wise prediction optimizes local decisions rather than the output melody itself---a selected subset is not yet a standalone melody.

We model a melody skeleton as a length-controlled, order-preserving latent subsequence, followed by a deterministic rhythmic-closure step that turns selected note events into a self-consistent reduced melody. Based on this view, we introduce MeloBottleneck, a self-supervised framework that learns the bottleneck through re-ornamentation reconstruction, a frozen autoregressive melody prior, and ornament-invariant learning.

Experiments on three benchmarks support this formulation, showing that latent-subsequence learning transfers more robustly than pseudo-label imitation and benefits downstream retrieval. A pseudo-label note classifier is strongest on synthetic out-of-distribution ornament-to-skeleton data, but MeloBottleneck transfers better to zero-shot and cross-domain settings. The extracted skeletons also improve BM25-based fragment retrieval under ornamentation and corruption, increasing retrieval quality while reducing query time.

Online materials are at {\scriptsize{\url{github.com/m-july/Supplementary-Files-For-MelobottleNeck-arXiv-Submission}}}, including a demo web page with paired melody-skeleton playback and the codebase repository.

\section{Melody Skeleton Extraction: Task Formulation and Prior Work}\label{sec:task}

\subsection{Task Formulation}\label{subsec:task_definition}

\begin{figure}[!htbp]
  \centering
  \includegraphics[alt={Three formulations of melody skeleton extraction. Prior-driven reduction directly outputs a closed skeleton, note-wise prediction outputs per-note scores or labels on the original melody, and our method models the skeleton as a length-controlled latent subsequence followed by rhythmic closure into a self-consistent reduced melody.},width=0.80\linewidth]{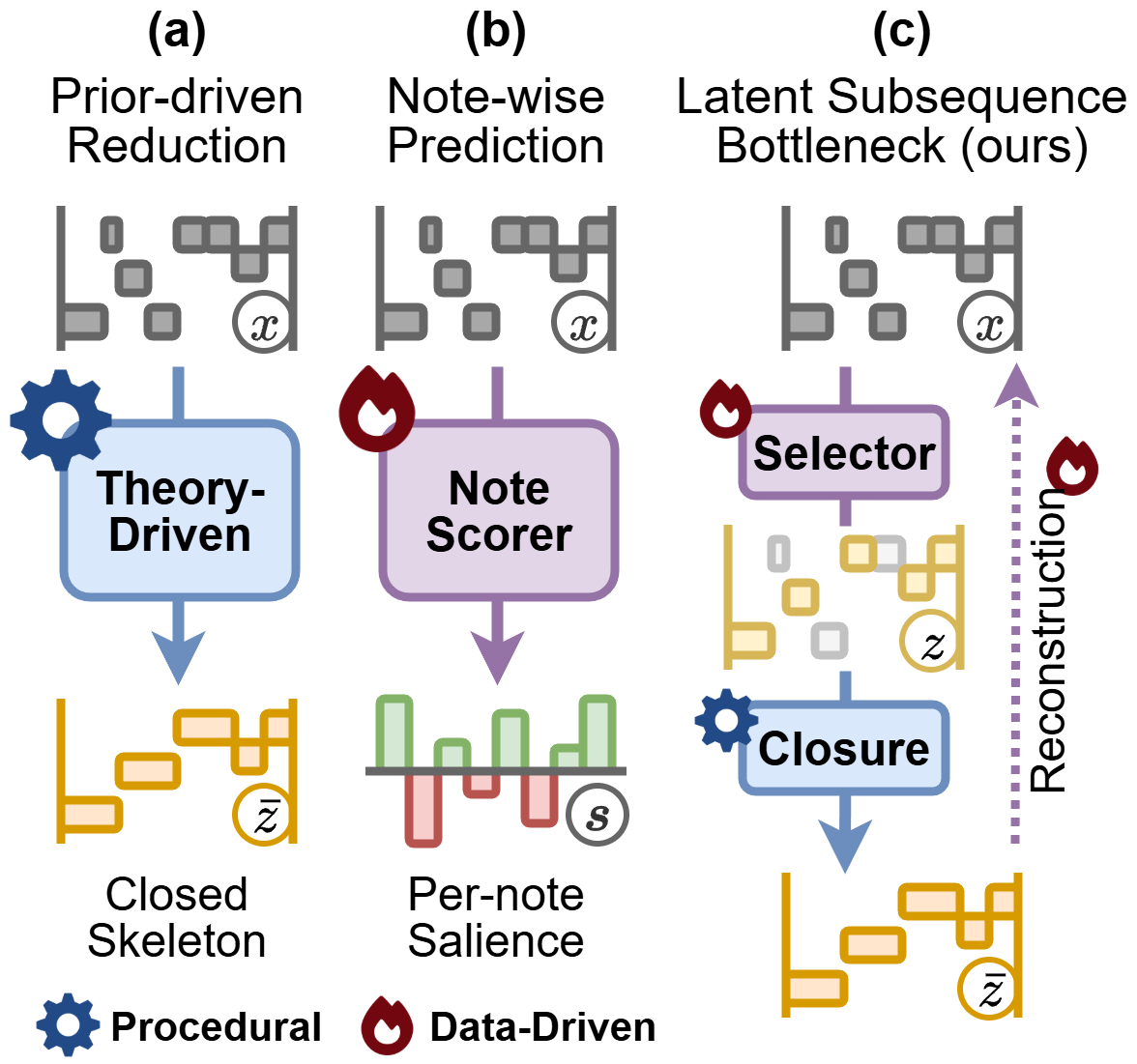}
  \vspace{-8pt}
  \caption{Three formulations of melody skeleton extraction.}
  \label{fig:method_comparison}
\end{figure}

Given a monophonic symbolic melody $x=(e_1,\ldots,e_L)$, melody skeleton extraction aims to derive a shorter melody that preserves structural note events while removing ornaments. We formulate the skeleton not as a set of independent note labels, but as an order-preserving latent subsequence $z=(e_{i_1},\ldots,e_{i_K})$, where $1 \le i_1 < \cdots < i_K \le L$ and $K$ is either predicted or specified by a target compression ratio $\rho$. \figref{fig:method_comparison} contrasts this view with prior-driven reduction and note-wise salience prediction.

Crucially, $z$ is not yet a standalone melody: deleting interior events changes the durations and gaps of retained notes. We therefore separate note selection from rhythmic closure, where selection yields $z$ and closure converts it into a standalone reduced melody $\bar z$. This places length control and melody-level coherence inside the task definition rather than in post-hoc heuristics.

\subsection{Relation to Prior Work}\label{subsec:task_prior_works}

Melodic reduction has long been studied through music-theoretic formalisms. Schenkerian analysis \cite{schenker1979free} and GTTM \cite{Book83Lerdahl:01} offer hierarchical views of structural importance. Early computational work by Frankel et al. \cite{CH76Frankel:01, IJMMS78Frankel:01} encoded Schenkerian elaboration procedures in LISP. Later systems implemented GTTM-style analyzers \cite{JNMR06Hamanaka:01} and probabilistic time-span trees \cite{CMMR15Hamanaka:01}. Previous works \cite{IJCAI07Gilbert:01, ISMIR16Groves:01} further used PCFGs for melodic reduction, Marsden \cite{ISMIR10Marsden:01} applied automatic Schenkerian reduction to variation recognition, and Wang et al. \cite{ISMIR25Wang:01} recently formulated reduction as shortest-path search. These methods are musically interpretable, but they typically encode strong analytical priors and are often tied to tonal/classical reduction formalisms.

Other work moves closer to note-wise structural prediction or downstream use of skeleton signals. Hu and Arthur \cite{FDMC21Hu:01} predict chord tones from features such as metric position, duration, and melodic intervals. Sun et al. \cite{ICMEW22Sun:01} use melodic skeletons for harmonization, and Zhang et al. \cite{ArXiv23Zhang:01} use skeleton-guided generation in WuYun. These directions show the utility of skeletal representations, but they either optimize local note decisions or treat skeletons primarily as auxiliary control signals, rather than as length-controlled reduced melodies that must themselves be coherent.

From a modeling perspective, our formulation is closer to discrete latent compression than to note classification. Previous works such as Extractive Summary as Discrete Latent Variables \cite{ArXiv18Komatsuzaki:01} and SEQ$^3$ \cite{NAACL19Baziotis:01} learn shorter latent sequences with explicit length control and language-model priors. BART \cite{ACL20Lewis:01} provides the denoising seq2seq pretraining paradigm; on the music side, MusicBERT \cite{ArXiv21Zeng:01} and PianoBART \cite{ICME24Liang:01} demonstrate the value of large-scale symbolic pretraining. MeloBottleneck combines these two strands in melody reduction: it learns an order-preserving latent note subsequence, turns it into a standalone melody through rhythmic closure, and optimizes it with reconstruction, melody-prior, and ornament-invariant objectives.

\section{Method: MeloBottleneck}\label{sec:method}

\begin{figure*}[!tbp]
  \centering
  \includegraphics[alt={},width=1.0\linewidth]{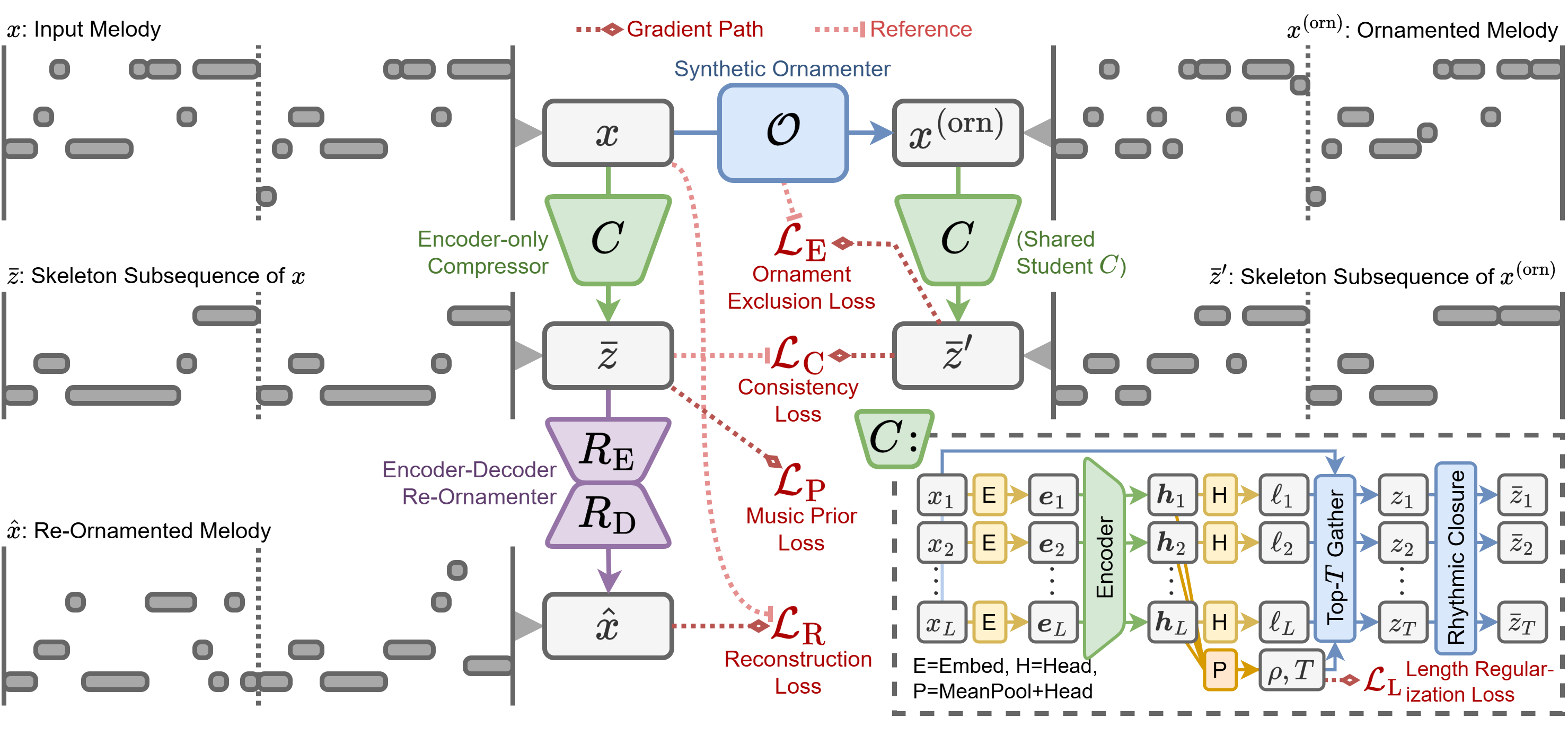}
  \vspace{-8pt}
  \caption{Overview of MeloBottleneck.}
  \label{fig:method_main}
\end{figure*}

Following Section~\ref{sec:task}, MeloBottleneck treats a melody skeleton as a shorter, order-preserving latent subsequence rather than a set of independent note labels. Given an augmented melody $x'=\mathcal A(x)$, the model selects a length-controlled subsequence $z$, converts it into a rhythmically closed skeleton $\bar z$, and trains the bottleneck by reconstructing $x'$ while remaining stable under procedural ornamentation (Figure~\ref{fig:method_main}).

\subsection{SimpleMono Event Representation}\label{subsec:representation}

We use a compact monophonic multi-attribute representation, \emph{SimpleMono}. After quantization at resolution $R=12$ positions per quarter note, each note with index $l$, onset $o_l$, and offset $f_l$, is encoded as a triple of (i) MIDI pitch $p_l \in [0, 127]$, (ii) duration $c_l^{D}$ and (iii) delta-time $c_l^{\Delta}$:
\begin{equation}
e_l=(p_l,c_l^{D},c_l^{\Delta}),
\end{equation}
where
\begin{equation}
c_l^{D}
=
\operatorname{clip}(f_l-o_l,\,1,\,N_D-1),
\end{equation}
\vspace{-10pt}
\begin{equation}
c_l^{\Delta}
=
\begin{cases}
\operatorname{clip}(o_{l+1}-f_l,\,-N_\Delta,\,N_\Delta-1), & l<L,\\
0, & l=L.
\end{cases}
\end{equation}
where $\operatorname{clip}(x,a,b)=\min(\max(x,a),b)$. A BOS event reuses the delta-time channel to encode a clipped onset anchor for the first note. We use SimpleMono sequences of up to 514 events, with $N_D=N_\Delta=96$. Vocab size is 421 (with 5 special tokens).

\begin{figure}[!htbp]
  \centering
  \includegraphics[alt={},width=1.0\linewidth]{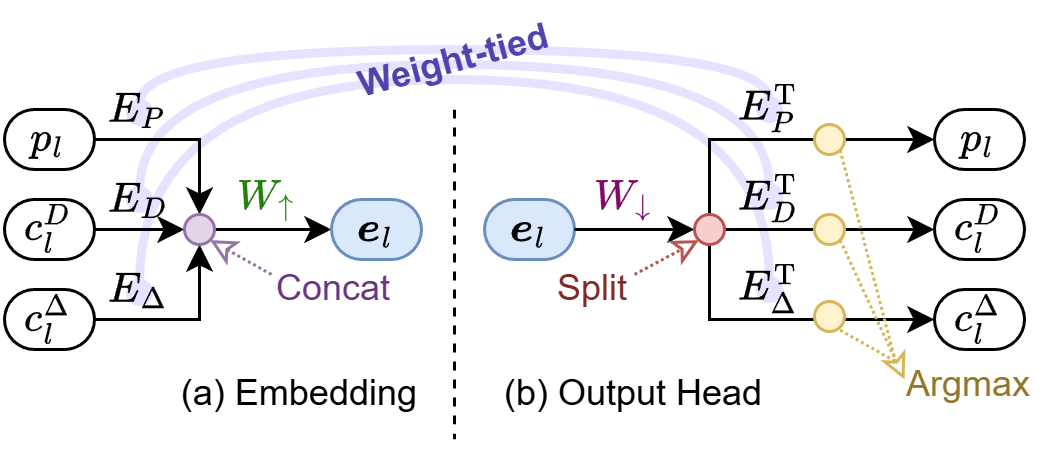}
  \vspace{-8pt}
  \caption{Multi-attribute representation.}
  \label{fig:embed_head}
\end{figure}

Our input embedding follows the concatenate-and-project construction used in OctupleMIDI-based compound-token models \cite{ArXiv21Zeng:01, ICME24Liang:01}, while SimpleMono reduces the event to a compact monophonic set of pitch, duration, and delta-time attributes and ties each attribute-specific output head to its corresponding input embedding table (see \figref{fig:embed_head}; $d_{\mathrm{attr\_embed}}=256$ and $d_{\mathrm{model}}=512$).

\subsection{Length-Controlled Subsequence Compressor}\label{subsec:compressor}

A Transformer encoder maps $x'=(e_1,\ldots,e_L)$ to contextual states $H=(h_1,\ldots,h_L)$. A selection head produces a logit $\ell_l$ for each note event and a note-selection distribution
\begin{equation}
s_l(x')
=
\frac{\exp(\ell_l)}{\sum_{j=1}^{L}\exp(\ell_j)} .
\end{equation}
A separate predictor estimates the compression ratio
\begin{equation}
\rho(x')
=
\rho_{\min}
+
(\rho_{\max}-\rho_{\min})
\,\sigma\!\big(\mathrm{MLP}(\operatorname{Pool}(H))\big),
\end{equation}
which yields a continuous bottleneck length and a hard length,
\begin{equation}
T_{\mathrm{cont}}=L\rho(x'),
\qquad
K=\lceil T_{\mathrm{cont}}\rceil .
\end{equation}
The hard latent subsequence $z$ is obtained by taking the top-$K$ events of $x'$ by $\ell_l$ and restoring source order. This makes the bottleneck an order-preserving latent subsequence rather than a note-wise label mask.

For backpropagation, the top-$K$ gathering is regarded as $K$ selection steps, and at each step $t$ we form a row-stochastic soft gather
\begin{equation}
\mathbf z_t^{\mathrm{soft}}=\sum_l p_t(l)E_{\mathrm{ori}}(e_l), \quad
p_t(l)=\mathrm{softmax}\Big(\frac{\tilde\ell_{t,l}}{\tau}\Big),
\end{equation}
where $\tilde\ell_{t,l}$ is a masked logit over remaining candidates with a leaky penalty $\kappa$ on masked-out indices.

To propagate gradients through variable length, we apply a soft gate (which takes effect at Eq.~\eqref{eq:gather})
\begin{equation}
\gamma_t
=
\sigma\!\left(
\frac{T_{\mathrm{cont}}-(t-0.5)}{T_\rho}
\right),
\qquad
T_\rho=0.5.
\end{equation}

In our experiments, $\rho_{\min} = 1/3$, $\rho_{\max} = 1$, $\kappa=0.7$, and $\tau$ is linearly annealed from $1.5$ to $0.5$.

\subsubsection{Length Regularization}

We regularize the predicted ratio distribution at the batch level. Let $\rho^{\mathrm{eff}}=K/L$, and let $\rho^{\mathrm{eff}}_{(1)}\le\cdots\le\rho^{\mathrm{eff}}_{(B)}$ be the sorted effective ratios in a batch of size $B$. We define target quantiles
\begin{equation}
a_b=\frac{b-0.5}{B}, \, q_b=
\operatorname{clip}\!\big(\mu_L+\sigma_L\Phi^{-1}(a_b),\,\rho_{\min},\,\rho_{\max}\big),
\end{equation}
and Length Regularization Loss
\begin{equation}
\mathcal L_{\mathrm{L}}
=
\sum_{b=1}^{B}
\big(\rho^{\mathrm{eff}}_{(b)}-q_b\big)^2.
\end{equation}
It exposes the model to a broad range of effective $\rho$'s during training, improving adaptability to different $\rho$ targets.

\subsubsection{Timeline Alignment Regularization}

To avoid early collapse to a narrow local region (e.g., suffix selection), inspired by DCTTS \cite{tachibana2018guidedattn} diagonal guided-attention, we further apply a Timeline Alignment Loss 
\begin{equation}
\mathcal L_{\mathrm{T}}
=
\frac{1}{\sum_t \gamma_t}
\sum_t \gamma_t
\sum_l p_t(l)\,
\frac{(u_l-v_t)^2}{2\sigma_T^2},
\end{equation}
where $u_l=o_l/o_{\mathrm{EOS}}$ is normalized score time, $v_t=(t-1)/\max(K-1,1)$ is normalized bottleneck progress, and $o_{\mathrm{EOS}}$ is the end time of $x'$. $\mathcal L_{\mathrm{T}}$ is applied only during early training, with its coefficient $\lambda_{\mathrm{T}}$ annealed to zero.

In our experiments, we use $\mu_L = 2/3$, $\sigma_L = 0.2$, and $\sigma_T = 0.075$.

\subsection{Rhythmic Closure}\label{subsec:closure}

The raw subsequence $z$ is not yet a valid melody: deleting interior events changes the meaning of duration and delta-time. We therefore apply a deterministic rhythmic closure operator $\bar z=\operatorname{Close}(z)$ before any sequence-level objective. Let $o_{i_t}$ denote the source onset of the $t$-th retained event and let $o_{\mathrm{EOS}}$ be the end time of the source melody. Closure keeps pitch and reassigns timing by absorbing removed score time into the preceding retained event:
\begin{equation}
\bar p_t = p_{i_t},
\qquad
\bar d_t =
\begin{cases}
o_{i_{t+1}}-o_{i_t}, & t<K,\\
o_{\mathrm{EOS}}-o_{i_t}, & t=K,
\end{cases}
\end{equation}
\begin{equation}
\bar e_t
=
\big(
\bar p_t,\,
\operatorname{clip}(\bar d_t,1,N_D-1),\,
0
\big).
\end{equation}
The closed skeleton becomes a standalone melody with zero inter-event gaps between adjacent retained events (see \figref{fig:method_comparison}~(c) for an instance).

The forward pass uses the hard closed skeleton, while gradients flow through the soft ordered gather:
\begin{equation}
\mathbf z_t^{\mathrm{ST}}
= 
\operatorname{sg}\!\big(E(\bar e_t)-\mathbf z_t^{\mathrm{soft}}\big)
+
\mathbf z_t^{\mathrm{soft}}, \quad \tilde{\mathbf z}_t^{\mathrm{ST}}=\gamma_t\,\mathbf z_t^{\mathrm{ST}}
\label{eq:gather}
\end{equation}
where $\operatorname{sg}(\cdot)$ denotes stop-gradient.

\subsection{Re-Ornamentation Reconstruction}\label{subsec:reconstruction}

A seq2seq reconstructor is trained to recover the augmented input $x'$ from the bottleneck, turning extraction into a re-ornamentation problem: if the bottleneck discards structural material, reconstruction becomes difficult. The reconstructor shares the embeddings and encoder with the compressor and uses a Transformer decoder initialized from denoising pretraining. We inject the compression ratio $\rho$ to the bottleneck states by FiLM modulation \cite{AAAI18Perez:01},
\begin{equation}
\tilde{\mathbf z}_t
=
\bigl(1+\gamma(\rho)\bigr)\odot \mathbf z_t^{\mathrm{ST}}
+
\beta(\rho), \,\,
(\gamma(\rho),\beta(\rho))=\mathrm{MLP}(\rho).
\end{equation}
so that the decoder can interpret the same structural content under different reduction levels. Reconstruction Loss is a duration-weighted multi-attribute cross-entropy:
\begin{equation}
\mathcal L_{\mathrm{R}}
=
\frac{
\sum_{l=1}^{L}
w_l
\big(
\lambda_P^{\mathrm{R}}\mathrm{CE}_{P,l}
+
\lambda_D^{\mathrm{R}}\mathrm{CE}_{D,l}
+
\lambda_\Delta^{\mathrm{R}}\mathrm{CE}_{\Delta,l}
\big)
}{
\sum_{l=1}^{L} w_l
},
\end{equation}
where $w_l=f_l-o_l$ is the target note duration in quantized positions. During training, we gradually mask a fraction of decoder inputs ($0\% \to 80\%$) so that reconstruction depends on the bottleneck rather than only on teacher-forced autoregressive context.

\subsection{Frozen Melody Prior}\label{subsec:prior}

Reconstruction encourages informativeness but not necessarily melody-likeness. We therefore train a decoder-only melody model on augmented melodies and freeze it during bottleneck training. From the soft ordered gather, we induce attribute distributions at bottleneck step $t$:
\begin{equation}
r_t^a(v)
=
\sum_{l=1}^{L}
p_t(l)\,
\mathbb I\!\big[a(e_l)=v\big],
\quad
a\in\{p,d,\Delta\}.
\end{equation}
Let $p_{\mathrm{LM}}^a(\cdot\mid \bar z_{<t})$ denote the frozen prior's next-attribute distribution given the hard closed prefix. Melody Prior Loss is
\begin{equation}
\mathcal L_{\mathrm{P}}
=
\sum_t
\sum_{a\in\{p,d,\Delta\}}
\alpha_a\,
\mathrm{KL}\!\big(
r_t^a
\,\|\, 
p_{\mathrm{LM}}^a(\cdot\mid \bar z_{<t})
\big).
\end{equation}
Because this prior is trained on ordinary melodies rather than on skeleton annotations, it acts as a generic melody-likeness constraint instead of a reduction teacher.

\subsection{Ornament-Invariant Learning}\label{subsec:invariance}

To make extraction stable under surface elaboration, we generate a strong view by procedurally ornamenting the weak view (details on ornamenter ${\mathcal{O}}$ at Section~4.4):
\begin{equation}
x'
\xrightarrow{\mathcal O}
\big(x^{\mathrm{orn}},\pi\big),
\end{equation}
where $\pi(k)\in\{1,\ldots,L\}\cup\{\varnothing\}$ maps the $k$-th ornamented event back to a source event, and $\pi(k)=\varnothing$ marks a newly inserted ornament. A teacher branch extracts $s(x')$ on the weak view in evaluation mode; a student branch processes $x^{\mathrm{orn}}$ under the same compression ratio budget $\rho(x')$ and outputs $s(x^{\mathrm{orn}})$. We align the student distribution back to the weak-view timeline by
\begin{equation}
\tilde s_l(x^{\mathrm{orn}})
=
\sum_{k:\pi(k)=l}
s_k(x^{\mathrm{orn}}), \, \, 
\hat s_l(x^{\mathrm{orn}})
=
\frac{\tilde s_l(x^{\mathrm{orn}})}
{\sum_j \tilde s_j(x^{\mathrm{orn}})}.
\end{equation}
We then optimize Ornament-Invariant Consistency Loss $\mathcal L_{\mathrm{C}}$ and Ornament Exclusion Loss $\mathcal L_{\mathrm{E}}$:
\begin{equation}
\mathcal L_{\mathrm{C}}
=
\mathrm{KL}\!\big(
s(x')
\,\|\,
\hat s(x^{\mathrm{orn}})
\big), \, \, 
\mathcal L_{\mathrm{E}}
=
\sum_{k:\pi(k)=\varnothing}
s_k(x^{\mathrm{orn}}).
\end{equation}
$\mathcal L_{\mathrm{C}}$ enforces ornament-invariant selection over source events, while $\mathcal L_{\mathrm{E}}$ explicitly suppresses mass assigned to synthetic insertions.

\subsection{Training Procedure}\label{subsec:training}

Training proceeds in three stages. (1) \emph{Denoising Pretraining} trains an encoder--decoder backbone on 
\begin{equation}
x
\xrightarrow{\mathcal A}
x'
\xrightarrow{\mathcal C}
\tilde x,
\end{equation}
where a BART-style corruption ${\mathcal C}$ \cite{ACL20Lewis:01, ICME24Liang:01} is adopted, to reconstruct $x'$ with
\begin{equation}
\mathcal L_{1}
=
\lambda_p^{\mathrm{A}}\mathrm{CE}_{p}
+
\lambda_d^{\mathrm{A}}\mathrm{CE}_{d}
+
\lambda_\Delta^{\mathrm{A}}\mathrm{CE}_{\Delta}.
\end{equation}
(2) \emph{Melody Prior Training} initializes a decoder-only language model from the pretrained decoder and trains it autoregressively on
\begin{equation}
x
\xrightarrow{\mathcal A_{\mathrm{LM}}}
x'
\end{equation}
with
\begin{equation}
\mathcal L_{2}
=
\lambda_p^{\mathrm{LM}}\mathrm{CE}_{p}
+
\lambda_d^{\mathrm{LM}}\mathrm{CE}_{d}
+
\lambda_\Delta^{\mathrm{LM}}\mathrm{CE}_{\Delta}.
\end{equation}
Finally, (3) \emph{Subsequence Bottleneck Training} initializes the compressor and reconstructor from Stage~1, freezes the Stage~2 prior, and optimizes
\begin{equation}
\mathcal J
=
\lambda_{\mathrm{R}}\mathcal L_{\mathrm{R}}
+
\lambda_{\mathrm{P}}\mathcal L_{\mathrm{P}}
+
\lambda_{\mathrm{L}}\mathcal L_{\mathrm{L}}
+
\lambda_{\mathrm{T}}\mathcal L_{\mathrm{T}}
+
\lambda_{\mathrm{C}}\mathcal L_{\mathrm{C}}
+
\lambda_{\mathrm{E}}\mathcal L_{\mathrm{E}}.
\end{equation}
At inference time, the extracted skeleton is the hard, rhythmically closed sequence $\bar z$; the compression ratio may be predicted by the model or externally specified. Exact numerical hyperparameters are summarized in Section~4.4.

\section{Experimental Setup}\label{sec:exp}

\begin{table*}[!btp]
  \centering
  \small
  \setlength{\tabcolsep}{4.2pt}

    \newcommand{\plaincell}[1]{%
      \shortstack[c]{\rule{0pt}{2ex}$#1$\\{\scriptsize \vphantom{($\pm 0.0000$)}}}%
    }
  
    \newcommand{\bestplaincell}[1]{%
      \shortstack[c]{\rule{0pt}{2ex}$\textbf{#1}$\\{\scriptsize \vphantom{($\pm 0.0000$)}}}%
    }
  
    \newcommand{\cellscore}[2]{%
      \shortstack[c]{\rule{0pt}{2ex}$#1$\\{\scriptsize ($\pm #2$)}}%
    }
  
    \newcommand{\bestcellscore}[2]{%
      \shortstack[c]{\rule{0pt}{2ex}$\textbf{#1}$\\{\scriptsize ($\pm #2$)}}%
    }

  \begin{tabular}{l|ccc|ccc|cc}
    \hline
    & \multicolumn{3}{c|}{O2B (OOD synth)} &
      \multicolumn{3}{c|}{V2T (cross-domain)} &
      \multicolumn{2}{c}{O2G (in-domain)} \\
    \cline{2-9}
    \shortstack[c]{Method\\{}} & Hard F1 $\uparrow$ & CFA $\uparrow$ & IM $\downarrow$ & Hard F1 $\uparrow$ & CFA $\uparrow$ & IM $\downarrow$ & Hard F1 $\uparrow$ & IM $\downarrow$ \\
    \hline
    \shortstack[c]{Random\\{}}
      & \cellscore{0.4703}{0.0011} & \cellscore{0.5335}{0.0008} & \cellscore{0.5221}{0.0011}
      & \plaincell{0.6138} & \plaincell{0.6156} & \plaincell{0.3653}
      & \plaincell{0.8058} & \plaincell{0.1928} \\
    \shortstack[c]{Uniform-Time\\{}}
      & \cellscore{0.6076}{0.0020} & \cellscore{0.6021}{0.0012} & \cellscore{0.3862}{0.0020}
      & \plaincell{0.6401} & \plaincell{0.6308} & \plaincell{0.3397}
      & \plaincell{0.8570} & \plaincell{0.1396} \\
    \shortstack[c]{Top-K Duration\\{}}
      & \cellscore{0.7865}{0.0017} & \cellscore{0.7300}{0.0008} & \cellscore{0.2964}{0.0015}
      & \plaincell{0.6508} & \plaincell{0.6402} & \plaincell{0.3271}
      & \bestplaincell{0.9317} & \plaincell{0.0641} \\
    \hline
    \shortstack[l]{\rule{0pt}{2ex}AMR-No-Harmony\\(heuristic)}
      & \cellscore{0.6841}{0.0025} & \cellscore{0.6417}{0.0014} & \cellscore{0.3114}{0.0024}
      & \plaincell{0.6267} & \plaincell{0.6226} & \plaincell{0.3532}
      & \plaincell{0.8797} & \plaincell{0.1194} \\
    \hline
    \shortstack[l]{\rule{0pt}{2ex}O2B-Learner\\(pseudo-label)}
      & \bestcellscore{0.8942}{0.0052} & \bestcellscore{0.7737}{0.0014} & \cellscore{0.1637}{0.0113}
      & \cellscore{0.6330}{0.0022} & \cellscore{0.6288}{0.0012} & \cellscore{0.3850}{0.0006}
      & \cellscore{0.8893}{0.0061} & \cellscore{0.1660}{0.0043} \\
    \hline
    \shortstack[l]{\rule{0pt}{2ex}MeloBottleneck\\(ours)}
      & \cellscore{0.8809}{0.0066} & \cellscore{0.7713}{0.0023} & \bestcellscore{0.0632}{0.0085}
      & \bestcellscore{0.6677}{0.0078} & \bestcellscore{0.6456}{0.0047} & \bestcellscore{0.3242}{0.0187}
      & \cellscore{0.8942}{0.0116} & \bestcellscore{0.0619}{0.0090} \\
    \hline
  \end{tabular}
  \vspace{-8pt}
  \caption{Main results under oracle compression ratio. Best results are in bold. Lower IM is better. O2B metrics and learning-based models are reported as mean{\scriptsize($\pm$std)} over 5 seeds (for dataset splitting and training inits).}
  \label{tab:main_results}
\end{table*}

\subsection{Data and benchmarks}
We train on a monophonic symbolic corpus aggregated from seven folk-song collections spanning Chinese, British, Irish, Swedish, Dutch, and worldwide repertoires \cite{ArXiv25Bu:01, jiugong_data1, BFDB25Deedman:01, Essen95Schaffrath:01, NorbeckABC26:01, Wu23TunesFormer:01, MTC19vanKranenburg:01}. File-level splitting (train:valid:test $=18{:}1{:}1$) and sliding-window segmentation are adopted. We evaluate on three benchmarks: (a) Main O2B, a synthetic OOD ornament-to-backbone task with unseen ornament operations and wider parameter ranges at test time; (b) TAVERN V2T, a zero-shot cross-domain variation-to-theme task obtained by aligning right-hand variations to their themes \cite{ISMIR15Devaney:01}; and (c) Jiugong O2G, a zero-shot in-domain benchmark pairing ornamented modern transcriptions \cite{Wang09Jiugong} (not provided in training data) with aligned \emph{gongche} skeletons from held-out volumes \cite{jiugong_data2}. See \tabref{tab:data_stats} for dataset statistics. Unless otherwise stated, evaluation uses the oracle ratio $\rho=K^\star/L$, so results isolate note selection rather than length prediction.

\begin{table}[!htbp]
  \centering
  \begin{tabular}{|l|c|c|c|c|}
    \hline
    \multirow{3}{*}{} & \multirow{2}{*}{train} & \multicolumn{3}{c|}{benchmark} \\
    \cline{3-5}
     &  & synth & \multicolumn{2}{c|}{real} \\
    \cline{2-5}
     & Main* & O2B* & V2T & O2G \\
    \hline
    \#sequence   & 47.1K   & 2.6K   & 619   & 20   \\
    \hline
    \#note       & 4.8M    & 269.7K & 12.7K & 3.1K \\
    \hline
    Time@80BPM   & 30d9h   & 1d17h  & 2h03m & 40m  \\
    \hline
  \end{tabular}
  \parbox{0.95\linewidth}{\footnotesize \textit{*: averaged over 5 random splits with different seeds.}}
  \vspace{-8pt}
  \caption{Dataset statistics for train set and benchmarks.}
  \label{tab:data_stats}
\end{table}

\subsection{Baselines}
We compare against three groups of baselines: (a) Simple ratio-controlled reducers---Random, Uniform-Time, and Top-K Duration; (b) AMR-No-Harmony, an adaptation of heuristic shortest-path automatic reduction \cite{ISMIR25Wang:01} without harmony but with explicit length control; and (c) O2B-Learner, a pseudo-label note classifier trained on procedural keep/delete labels from the same online ornamenter ${\mathcal O}$ used in ornament-invariant training. O2B-Learner uses the same SimpleMono representation and encoder capacity as our compressor, and its scores are converted to skeletons by top-$K$ selection under the same ratio budget.

\subsection{Metrics}
Because benchmark references are aligned retained-event subsequences rather than independently annotated closed reductions, quantitative evaluation focuses on note-event selection; models are run with the reference-length budget, so these metrics isolate selection quality rather than free length prediction. We report (i) Hard F1 between the predicted hard selection and the aligned reference events, (ii) normalized Cut-F1 AUC (CFA), which computes top-$k$ F1 over cut ratios $r \in \{m/12\}_{m=4}^{12}$ and normalizes the trapezoidal AUC by the cut-range length, and (iii) Insertion Mass (IM), which is the normalized selection mass assigned to reference inserted/non-skeleton input events. CFA is omitted for O2G because its reference compression ratios have limited spread, making CFA on O2G less informative. All selection metrics are computed over non-special note events and macro-averaged over pieces. Strong-beat Lift and Duration Lift are also reported as diagnostic musical-bias metrics computed from selection-score distributions.

\subsection{Implementation Details}
The backbone has 6 encoder layers, 3 decoder layers, and 8 heads, totaling 33.2M parameters; O2B-Learner adopts the same encoder and Stage~1 initialization. Training follows Section~\ref{subsec:training}, with 160/80/1 epochs for \mbox{Stages~1--3}. ${\mathcal{A}}$ and ${\mathcal{A}_{\mathrm{LM}}}$ use transposition and time scaling, the latter favoring duration lengthening. $\mathcal{O}$ comprises pre-grace, post-grace, between-insert, trill, and rearticulation; $\mathcal{O}_{\mathrm{OOD}}$ further uses out-of-distribution parameters (e.g., wider sampling range for ornamental pitch jitter) and adds turn and pair-repeat. $\mathcal{C}$ applies masking, deletion, and document rotation. We set $(\lambda_{\mathrm{R}}, \lambda_{\mathrm{P}}, \lambda_{\mathrm{L}}, \lambda_{\mathrm{C}}, \lambda_{\mathrm{E}})=(1.8,0.6,10.0,4.0,2.0)$ and anneal $\lambda_{\mathrm{T}}$ from 0.1 to 0. Additional details (e.g., dataset details and efficiency statistics) are provided in the online material mentioned in Section~1.

\vspace{-5pt}

\section{Results}\label{sec:results}

\begin{table*}[!tbp]
  \centering
  \small
  \setlength{\tabcolsep}{4.0pt}

  \newcommand{\cellscore}[2]{%
    $#1$\scriptsize($\pm #2$)
  }

  \begin{tabular}{l|c|c|c|cc}
    \hline
    & O2B (OOD synth) &
      V2T (cross-domain) &
      O2G (in-domain) &
      \multicolumn{2}{c}{Diagnostic lifts on O2B} \\
    \cline{2-6}
    & Hard F1 & Hard F1 & Hard F1 & Strong-beat & Duration \\
    \hline
    Full & \cellscore{0.8809}{0.0066} & \cellscore{0.6677}{0.0078} & \cellscore{0.8942}{0.0116}
         & \cellscore{1.8342}{0.0744} & \cellscore{1.7878}{0.1122} \\
    \hline
    w/o $\mathcal L_{\mathrm{R}}$ & \cellscore{-0.0034}{0.0164} & \cellscore{-0.0153}{0.0140} & \cellscore{-0.0224}{0.0143}
      & \cellscore{-0.0740}{0.0883} & \cellscore{-0.2336}{0.1087} \\
    w/o $\mathcal L_{\mathrm{P}}$ & \cellscore{-0.0546}{0.0105} & \cellscore{-0.0051}{0.0084} & \cellscore{-0.0003}{0.0130}
      & \cellscore{+0.2122}{0.1124} & \cellscore{+2.0316}{0.2222} \\
    w/o $\mathcal L_{\mathrm{C}}$ & \cellscore{-0.1007}{0.0167} & \cellscore{-0.0071}{0.0052} & \cellscore{-0.0359}{0.0084}
      & \cellscore{+0.4696}{0.2187} & \cellscore{+2.3450}{0.4068} \\
    w/o $\mathcal L_{\mathrm{E}}$ & \cellscore{-0.1859}{0.0265} & \cellscore{-0.0011}{0.0052} & \cellscore{-0.0143}{0.0082}
      & \cellscore{-0.4306}{0.1549} & \cellscore{-0.2052}{0.2158} \\
    w/o Rhythmic Closure & \cellscore{-0.0032}{0.0076} & \cellscore{-0.0053}{0.0109} & \cellscore{-0.0030}{0.0095}
      & \cellscore{-0.1000}{0.1333} & \cellscore{-0.2266}{0.1330} \\
    \hline
  \end{tabular}
  \vspace{-8pt}
  \caption{Ablations of MeloBottleneck. Rows after \textit{Full} report changes ($\Delta$) from the full model.}
  \label{tab:ablation}
  \vspace{-5pt}
\end{table*}

\subsection{Main comparison}
Table~\ref{tab:main_results} reports Hard F1, CFA, and IM under the oracle ratio. On synthetic O2B, the pseudo-label classifier achieves the best Hard F1 and CFA, even under OOD ornamentation, which is consistent with its objective of directly imitating procedural keep/delete labels. However, on the two zero-shot transfer benchmarks, V2T and O2G, MeloBottleneck is stronger: it outperformed both the heuristic and the pseudo-label baselines in all 3 metrics. This supports our central claim: latent-subsequence learning transfers more robustly than note-wise pseudo-label imitation. Among non-learning baselines, Top-K Duration is particularly strong on O2G, suggesting that the Jiugong references correlate with long notes, but it is clearly weaker on O2B and V2T.

\vspace{-5pt}

\subsection{Ablations and bias diagnostics}

Table~\ref{tab:ablation} isolates the main training signals. Removing Ornament Exclusion causes the largest degradation on O2B and sharply increases IM, confirming that explicit suppression of synthetic insertions is critical for ornament recovery. Removing Ornament-Invariant Consistency or the Melody Prior reduces transfer performance. Removing Reconstruction mainly hurts V2T and O2G, indicating that re-ornamentation provides a global training signal beyond local pseudo-label imitation. Duration lift doubled without $\mathcal{L}_{\text{P}}$ or $\mathcal{L}_{\text{C}}$, suggesting that these objectives prevent the selector from collapsing to simple duration heuristics.

\vspace{-5pt}

\subsection{Skeletonization improves robust retrieval}

\begin{table}[!bp]
  \vspace{-5pt}
  \centering
  \small
  \setlength{\tabcolsep}{3.6pt}
  \begin{tabular}{l|cc|c|c}
    \hline
    Setting & R@1 & R@10 & MRR & Time/query \\
    \hline
    Random (219 docs) & 0.0046 & 0.0457 & 0.0205 & -- \\
    Full-seq BM25 & 0.1170 & 0.2442 & 0.1592 & 0.367ms \\
    Skeleton BM25 (ours) & \textbf{0.2084} & \textbf{0.3579} & \textbf{0.2584} & \textbf{0.274ms} \\
    \hline
  \end{tabular}
  \vspace{-5pt}
  \caption{Results of BM25 fragment retrieval under OOD ornamentation and transcription-like corruptions.}
  \label{tab:retrieval}
  \vspace{-2pt}
\end{table}

We further evaluate whether extracted skeletons benefit a downstream task: fragment-to-document retrieval using sparse BM25 matching. We construct 10K ornamented and corrupted query fragments (with $8 \sim 64$ notes) from held-out Jiugong volumes and retrieve from a collection of 219 documents using hashed 3-grams of relative pitch and duration transitions. In this experiment, we run MeloBottleneck in inference mode and skeletonize \emph{both} queries and documents, with predicted compression ratios ($\rho$'s) for queries and fixed ratio $\rho = 0.8$ for documents (to prevent over-simplification on documents). Table~\ref{tab:retrieval} shows that skeletonization substantially improves retrieval quality while reducing per-query time, consistent with the hypothesis that skeletons suppress ornament-induced n-gram mismatches and shorten sequences for more efficient indexing and scoring.

\vspace{-6pt}

\section{Discussion and Limitations}\label{sec:discussion}

\vspace{-3pt}

\noindent\textbf{Notion of skeleton differs.}
Melody skeletons are not uniquely defined under different analytical objectives and downstream tasks. MeloBottleneck learns an \emph{operational} skeleton induced by our objectives---an order-preserving latent subsequence that (i) retains information for re-ornamentation reconstruction, (ii) forms a standalone melody, and (iii) remains stable under ornamentation.

\noindent\textbf{Dependence on procedural ornamentation.}
Although MeloBottleneck avoids using keep/delete pseudo-labels as explicit training targets, it still relies on a procedural ornamenter to generate strong views and to identify inserted events. If this ornamenter mismatches a target style, the invariance signal may under-penalize real ornaments or over-penalize stylistic surface notes.

\noindent\textbf{Evaluation caveats.}
We evaluate selection under an oracle compression ratio; this leaves free-length prediction only indirectly tested (except in retrieval, where $\rho$ is predicted). V2T targets are alignment-induced correspondences between variations and themes, rather than human reduction. An independent evaluation of the musical quality of the closed skeletons is not provided.

\noindent\textbf{Scope.}
We focus on monophonic symbolic melodies and a simple rhythmic-closure operator that absorbs deleted score time into the preceding retained event, yielding zero inter-event gaps. Extending MeloBottleneck to polyphony and harmony-aware reduction would require richer representations and closure operators.

\vspace{-6pt}

\section{Conclusion}\label{sec:conclusion}

\vspace{-3pt}

We introduced MeloBottleneck, a self-supervised framework that models a melody skeleton as a length-controlled, order-preserving latent subsequence rather than a set of independent note-wise labels. A deterministic rhythmic closure operator turns the selected subsequence into a self-consistent reduced melody. Training combines re-ornamentation reconstruction, a frozen autoregressive melody prior, and ornament-invariant objectives. Experiments across synthetic OOD ornament-to-skeleton, cross-domain variation-to-theme, and in-domain gongche benchmarks suggest a clear trade-off: pseudo-label imitation can be strongest when evaluation matches the generator, while latent-subsequence learning transfers more robustly and suppresses inserted ornaments more cleanly. The extracted skeletons also improve ornament-and-corruption-robust fragment retrieval, indicating practical value as a representation for downstream tasks.

\section{AI Usage Statement}\label{sec:ai_usage}

AI tools were used in this study to support technical exploration, including surveying candidate methods under given specific design requirements. All adopted methods were retained only after experimental validation and manual inspection. AI tools were also used to assist in architectural optimization and to suggest candidate hyperparameter settings, while the final architecture and hyperparameters were determined through validation-based ablation studies. Additional AI assistance was limited to programming and academic writing support. All references, datasets, and quantitative results were manually verified by the authors. The authors take full responsibility for all technical choices, experiments, and claims.

\begin{CJK*}{UTF8}{gbsn}
\bibliography{references}
\end{CJK*}

%
%
%
%

\end{document}